\documentclass[%
 reprint,
 amsmath,amssymb,
 aps,
]{revtex4-2}

\usepackage{algorithm}
\usepackage{algorithmic}
\usepackage{float}
\usepackage{subfig}
\usepackage{graphicx}
\usepackage{dcolumn}
\usepackage{bm}

\begin{document}

\preprint{APS/123-QED}

\title{A Hierarchical Fused Quantum Fuzzy Neural Network for Image Classification}

\author{Sheng-Yao Wu}
\affiliation{State Key Laboratory of Networking and Switching Technology, Beijing University of Posts and Telecommunications, Beijing, 100876, China}
\author{Run-Ze Li}
\affiliation{State Key Laboratory of Networking and Switching Technology, Beijing University of Posts and Telecommunications, Beijing, 100876, China}
\author{Yan-Qi Song}
\affiliation{State Key Laboratory of Networking and Switching Technology, Beijing University of Posts and Telecommunications, Beijing, 100876, China}
\author{Su-Juan Qin}
\email{qsujuan@bupt.edu.cn}
\affiliation{State Key Laboratory of Networking and Switching Technology, Beijing University of Posts and Telecommunications, Beijing, 100876, China}
\author{Qiao-Yan Wen}
\email{wqy@bupt.edu.cn}
\affiliation{State Key Laboratory of Networking and Switching Technology, Beijing University of Posts and Telecommunications, Beijing, 100876, China}
\author{Fei Gao}
\email{gaof@bupt.edu.cn}
\affiliation{State Key Laboratory of Networking and Switching Technology, Beijing University of Posts and Telecommunications, Beijing, 100876, China}

\date{\today}

\begin{abstract}
Neural network is a powerful learning paradigm for data feature learning in the era of big data. However, most neural network models are deterministic models that ignore the uncertainty of data. Fuzzy neural networks are proposed to address this problem. FDNN is a hierarchical deep neural network that derives information from both fuzzy and neural representations, the representations are then fused to form representation to be classified. FDNN perform well on uncertain data classification tasks. In this paper, we proposed a novel hierarchical fused quantum fuzzy neural network (HQFNN). Different from classical FDNN, HQFNN uses quantum neural networks to learn fuzzy membership functions in fuzzy neural network. We conducted simulated experiment on two types of datasets (Dirty-MNIST and 15-Scene), the results show that the proposed model can outperform several existing methods. In addition, we demonstrate the robustness of the proposed quantum circuit. 
\end{abstract}

\pacs{Valid PACS appear here}
\maketitle

\section{Introduction}

In the era of big data, a diverse and vast amount of data is generated. One concern is that the data contains unpredictable uncertainties\cite{seltzer2013investigation}. These uncertainties may arise from a variety of factors, including data ambiguity, incompleteness, noise, and redundancy. In the actual data analysis process, uncertainty may make it difficult for the model to capture the real patterns in the data, which brings great challenges to data classification tasks \cite{gawlikowski2023survey}.

Most of the existing learning models are deterministic algorithms, which face difficulties when dealing with uncertain data. To solve this problem, fuzzy learning was proposed and used in fields such as image processing\cite{bhalla2022fuzzy}], financial analysis\cite{lee2019chaotic}, and control systems\cite{hou2019finite}. Fuzzy systems can automatically learn fuzzy membership functions from a large amount of training data and derive fuzzy rules accordingly. Following the determination of fuzzy membership functions, fuzzy logic values for the data can be obtained. Subsequently, these fuzzy logic values are effectively defuzzied through linear combination to form features for the final classification task. Compared with traditional deterministic representation, the automated learning process of fuzzy logic representation enables it to effectively represent the uncertainty data.

In recent years, the development of quantum computing has provided new perspectives for data processing\cite{grover1998framework}. Quantum machine learning (QML) is a new research field that aims to take advantage of quantum computing to solve machine learning tasks\cite{biamonte2017quantum}. Quantum kernels have been used for support vector machines for classification problems\cite{blank2020quantum}. Variational quantum circuits are also used to design variational quantum classifiers \cite{huang2023near}. Quantum neural network (QNN) which are composed of variational quantum circuits is proposed to process data \cite{li2022recent}. It has been shown that encoding data into gate-based quantum circuits can generate kernels for machine learning models via quantum measurements. Some researchers have also studied the computational advantages of QML models\cite{liu2021rigorous}. Thus, QNNs have great potential in dealing with problems that are difficult to solve with classical neural networks\cite{perez2020data}. 

\begin{figure*}[!t]
\centering
\includegraphics[width=6in]{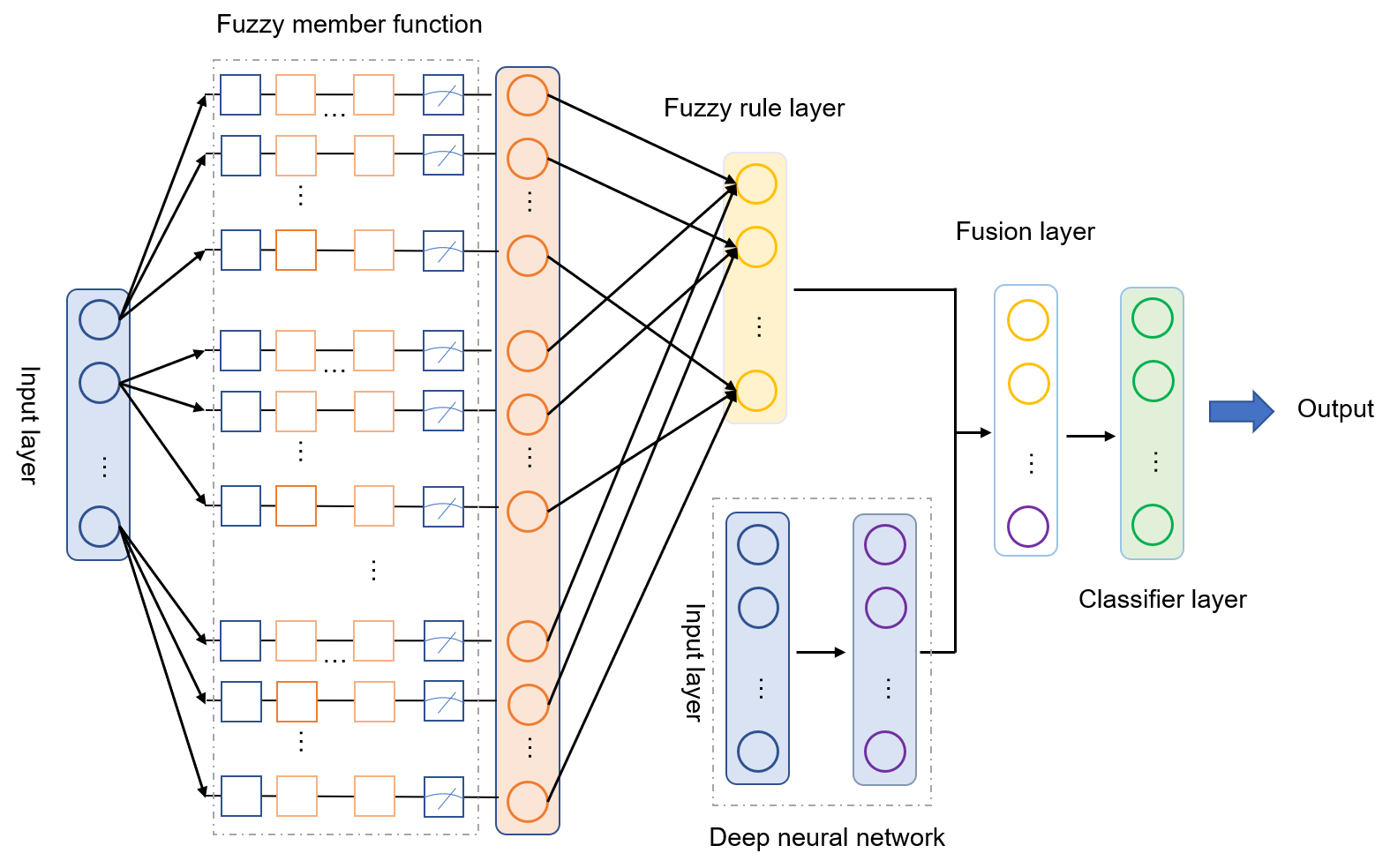}
\caption{The structure of HQFNN. It is composed of four parts that respectively address different learning tasks, i.e. quantum fuzzy logic representation, deep neural network representation, fusion layer and classifier layer.}
\label{fig_1}
\end{figure*}

Deng et al. proposed FDNN\cite{deng2016hierarchical}, which simultaneously extracts features from fuzzy and classical neural networks and fuses them through a fusion layer to form the final representation used for data classification. FDNN is verified to perform well on data classification tasks. 

In this paper, in order to investigate the performance of QNN in fuzzy neural networks, we proposes a novel hierarchical fused quantum fuzzy neural network (HQFNN). Different from FDNN, we use QNNs to learn fuzzy membership functions in fuzzy neural network. The proposed model is suitable for difficult tasks containing data ambiguity and noise. We verified the performance of the proposed model on two types of image datasets (Dirty-MNIST and 15-Scene). Experimental results show that the proposed model performs better than several existing methods. This research provides new ideas for the application of QNN in the field of data classification.

Note that a study has been proposed that combines QNN, classical networks and fuzzy logic to introduct quantum fuzzy neural network (QFNN) for sentiment and sarcasm detection\cite{tiwari2024quantum}. QFNN is a mltimodal fusion and multitask learning algorim with Sqe2Seq structure, and the model use QNN in deffuzzifier to obtain predictions. The method and objective of QFNN are both different from our model.

The contributions of this paper are as follows:

(1) A HQFNN network is proposed, QNNs are used to learn fuzzy membership functions in fuzzy neural network. Multi-modal learning is applied to fuse the fuzzy logic representation obtained by the QNNs and the neural representation obtained by the classical network to increase the accuracy in image classification tasks.

(2) Circuit structures of QNNs are designed to extract fuzzy logic representation from data. The proposed circuit structure utilizes only single qubit and consists of only a small number of single-qubit rotation gates, making it easy to implement.

(3) Experiments are performed on two types of image datasets to verify the performance of the proposed model. Compared with several existing classic networks, it can achieve better performance. The experiment simulated the performance of a single-qubit quantum circuit in a noisy environment, the results demonstrate that the proposed QNN exhibits strong robustness.

The rest of this paper is organized as follows. In Section \ref{section:sec2}, we discuss works related to HQFNN. We provide a detailed description of the proposed model in Section \ref{section:sec3}. Section \ref{section:sec4} outlines the experimental setup and presents experimental results to show the performance of the model. Section \ref{section:sec5} discusses the structures of quantum circuit and the methods for data encoding. Section \ref{section:sec6} concludes this paper and looks forward to potential future research directions.

\section{Related Work}\label{section:sec2}

\subsection{Image Classification}

Image classification is of great significance in the fields of computer vision and machine learning. Early image classification methods were based on hand-designed feature extraction algorithms, such as SIFT and HOG \cite{van2008kernel}, and then used support vector machines (SVM) or k nearest neighbors (KNN) for classification. This type of method only works well for simple images and small-scale datasets. As image scenes become complex and data size increases, this type of method is no longer applicable.

The emergence of CNN has made a major breakthrough in image classification tasks. CNN extracts the feature representation of the image through convolution and pooling operations. Subsequently, a large number of CNN-based models were proposed to improve the performance of image classification. In 1998, LeCun et al. \cite{lecun1998gradient} proposed LeNet. It was one of the early neural networks used for handwritten digit recognition, laying the foundation for later more complex deep learning models. LeNet is a multi-layer neural network, mainly composed of convolutional layers, pooling layers and fully connected layers. Its core idea is to gradually extract the features of the image through convolution and pooling operations, and map the extracted features to the output category through a fully connected layer. Krizhevsky et al. \cite{krizhevsky2012imagenet} proposed AlexNet in 2012, which is a deep convolutional neural network that uses more convolutional layers and parameters, and uses technologies including ReLU activation functions, data enhancement and dropout, which extremely improves the performance and generalization capabilities of the network. In 2014, Simonyan \cite{simonyan2014very} proposed VGGNet. Its core idea is to gradually extract the features of the image by stacking multiple small-sized convolution kernels and pooling layers, with a simple and unified network structure. It uses a parameter sharing strategy to reduce the number of parameters and improve training efficiency and generalization ability. In the same year, Szegedy et al. \cite{szegedy2015going} proposed a new deep CNN GoogLeNet. GoogLeNet introduces the inception module, which combines convolution kernels and pooling operations of different scales into the same module, thus increasing the width and depth of the network and improving the expressive ability of the network. An auxiliary classifier is introduced to alleviate the vanishing gradient problem and reduce overfitting. In 2015, He et al. \cite{he2016deep} proposed ResNet, which introduced the concept of residual learning, solved the problem of gradient disappearance and degradation in deep network training, and brought breakthroughs to the research of image classification and other computer vision tasks.

In recent years, image classification algorithms based on attention mechanisms have also appeared \cite{dai2021coatnet}. By learning the key areas in the image, the degree of attention to important areas is increased, thereby improving the classification performance. In addition, Howard et al. \cite{howard2017mobilenets} used transfer learning for image classification, using the trained model to fine-tune on new tasks, accelerating the training of the model and improving the generalization ability. However, the above methods all process deterministic image data. It cannot handle ambiguity and uncertainty data well.

\subsection{Classical Fuzzy Logic}

In 1965, Zadeh \cite{zadeh1965fuzzy} proposed the concept of fuzzy logic, which showed great advantages in solving practical problems. There are many models based on classic fuzzy logic. In 2019, Sarabakha et al.\cite{sarabakha2019online} used antecedent fuzzification to learn the control of nonlinear systems. The proposed fuzzy neuro network is a sequence structure. In the same year, Deng et al. \cite{deng2016hierarchical} constructed a fuzzy neural network FDNN which used fuzzy representation of data and neural network representation for information derivation, and achieved good performance in scene image classification and brain MRI segmentation tasks. In 2022, Wang et al. \cite{wang2023novel} proposed a fuzzy neural information processing block that can extract the input image into the fuzzy domain. Compared with existing super-resolution methods, the proposed method has better performance in high resolution medical image reconstruction, and reduces model parameters and analysis time. In 2023, Yazdinejad \cite{yazdinejad2023optimized} proposes an optimized fuzzy deep learning (OFDL) model for data classification based on non-dominated sorting genetic algorithm II. OFDL utilizes Pareto optimal solutions for multi-objective optimization using NSGA-II based on their objective functions to reach optimized backpropagation and fuzzy membership functions. The above models use classical neural networks without introducing QNNs.

\subsection{Quantum Machine Learning}

Quantum machine learning is an emerging research field that applies quantum computing to machine learning. Quantum machine learning includes quantum support vector machines (QSVM), quantum linear regression, quantum clustering and QNNs, etc. Quantum machine learning uses the parallelism of quantum computing to accelerate certain classical algorithms. In addition, QNNs use the properties of quantum entanglement and superposition to allow data to be better represented \cite{long2023enhanced}.

In 1995, Kak et al. \cite{kak1995quantum} proposed the concept of quantum neural computing, laying a theoretical foundation for quantum neural networks. In 2003, Anguita et al. \cite{anguita2003quantum} proposed a QSVM algorithm, which uses Grover search to accelerate the training process of the SVM data classification algorithm. In 2012, Wiebe et al. \cite{wiebe2012quantum} first proposed a quantum linear regression algorithm using the HHL algorithm. When the data matrix is a sparse matrix and has a very low condition number, this algorithm has exponential acceleration compared to the classical algorithm. In 2014, Lloyd et al. \cite{lloyd2013quantum} proposed a distance-based classification algorithm, which can achieve exponential acceleration compared with the classic algorithm. In 2019, Kerenidis et al. \cite{kerenidis2019q} proposed the q-means clustering algorithm. Compared to the classic k-means algorithm, the q-means algorithm provides exponential speedups on the number of points in the data set. However, the above algorithm needs to be implemented on a large-scale quantum computer that is tolerant to noise. In the era of noisy intermediate-scale quantum computing, variational quantum algorithms (VQA) have been proposed and widely studied. In 2018, Mitarai et al. \cite{mitarai2018quantum} proposed a QML framework based on parameterized quantum circuit (PQC), called quantum circuit learning (QCL), and proposed the idea of using unitary operator gates to approximate nonlinear functions. In 2018, Lloyd \cite{lloyd2018quantum} first proposed the concept of quantum generative adversarial network (QGAN) and analyzed the potential quantum advantages of QGAN from a theoretical perspective. In 2019, Cong et al. \cite{cong2019quantum} proposed a quantum convolutional neural network (QCNN) based on variational circuits, which constructed a convolution layer and a pooling layer with parameterized quantum gates, and proved it on two types of problems: quantum phase identification and quantum error correction. In 2020, Schuld et al. \cite{schuld2020circuit} proposed a low-depth parameterized quantum circuit as a binary classifier. In 2022, Qu et al. \cite{qu2022temporal} proposed a quantum graph convolutional neural network to solve the traffic congestion prediction problem. The results showed that the algorithm can effectively solve the problem. In 2023, Skolik et al. \cite{skolik2023robustness} studied the performance of variational quantum reinforcement learning under shot noise, coherent and incoherent error noise sources, and proposed methods to reduce the measurements required to train Q-learning agent. In general, the research on QML models has covered all aspects and has broad application prospects.

\subsection{Quantum Fuzzy Logic}

Quantum fuzzy logic can be seen as a combination of quantum theory and fuzzy logic. In 2007, Menichenko et al. \cite{melnichenko2007quantum} proposed that quantum logic can also be regarded as a kind of fuzzy logic and provided a theoretical proof. This lays the theoretical foundation for quantum fuzzy logic. In 2023, Tiwari et al. \cite{tiwari2024quantum} proposed a quantum fuzzy logic and applied it to multi-modal sentiment and sarcasm detection. Currently, there are few studies on quantum fuzzy logic, which is a field worthy of further exploration.

\section{Proposed Method}\label{section:sec3}
The HQFNN is composed of four main network parts: quantum fuzzy logic representation, deep neural network representation, fusion layer and classifier layer. The architecture of HQFNN is shown in Fig. \ref{fig_1}. The input data will be input into the quantum fuzzy logic representation network and the deep neural network to generate fuzzy logic representation and neural network representation of the data respectively. Then, the features of the two parts are fused through the aggregation module. Finally, the fused features are input to a classification neural network module for processing, and the classification neural network module classifies the data into different categories.

\begin{figure*}[!t]
\centering
\includegraphics[width=6in]{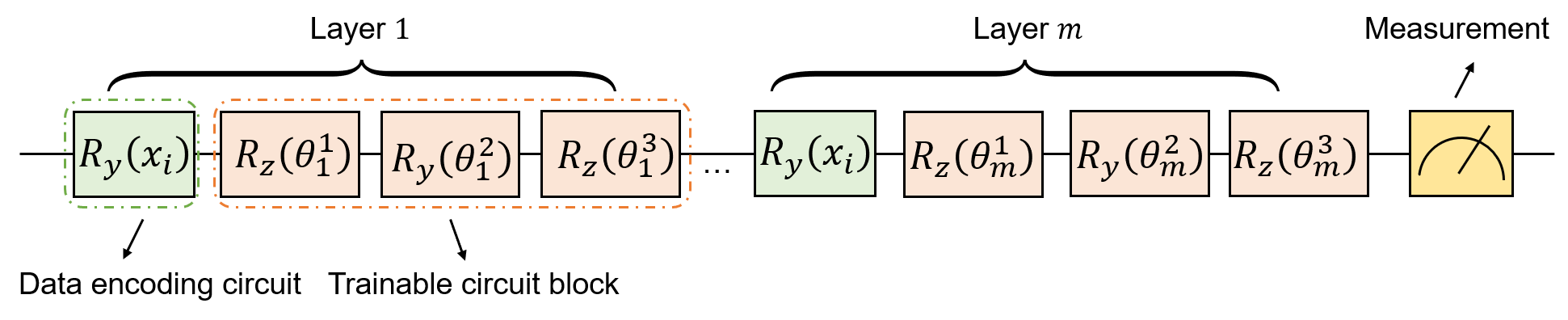}
\caption{Quantum circuit with single-qubit for quantum member function.}
\label{fig_2}
\end{figure*}

\textbf{Quantum fuzzy logic representation}. Each node of the input layer is connected to multiple membership functions, and each input variable is assigned a membership degree for different fuzzy sets. Generally speaking, an input variable is an element of the input vector. Fuzzy membership functions calculate the degree to which an input node belongs to a specific fuzzy set. For the $k$-th fuzzy membership function, this is a mapping ${u_k}:R \to \left[ {0,1} \right]$. In previous work \cite{lin2001self,lin2006support}, a Gaussian function with a mean of $\mu$ and a variance of ${\sigma ^2}$ is selected as the membership function. This paper proposes to use the QNN as shown in Fig. \ref{fig_2} as the membership function.

Fuzzy sets were first proposed by L.A. Zadeh in 1995 \cite{zadeh1965fuzzy}. Fuzzy sets are defined as

\begin{equation}
\label{eq1}
A = \left\{ {\left( {x,{\mu _A}\left( x \right)} \right)|x \in X} \right\}.
\end{equation}

In which, ${\mu _A}$ is the membership function of the fuzzy set $A$, that is, the target is mapped to its fuzzy representation. $X$ is the domain, representing the set of all input variables. In the representation of a QNN (usually implemented with a parametrized quantum circuit), the membership function can be expressed as

\begin{equation}
\label{eq2}
{\mu _A}\left( x \right) = {f_\theta }\left( x \right) = \left( {\left\langle 0 \right|{U^\dag }\left( {x,\theta } \right)MU\left( {x,\theta } \right)\left| 0 \right\rangle  + 1} \right)/2.
\end{equation}

Where $\left| 0 \right\rangle$ represents the initial state of the quantum computer, $U\left( {x,\theta } \right)$ is the quantum circuit that depends on $x$ and $\theta$, $x$ is the input and $\theta$ is an adjustable parameter. $M$ is an observable. ${f_\theta }\left( x \right)$ can be estimated by measuring the quantum circuit multiple times and aggregating the measurement outcomes. Since $\left\langle 0 \right|{U^\dag }\left( {x,\theta } \right)MU\left( {x,\theta } \right)\left| 0 \right\rangle $ is the expectation under the observable $M$, then $\left\langle 0 \right|{U^\dag }\left( {x,\theta } \right)MU\left( {x,\theta } \right)\left| 0 \right\rangle  \in \left[ { - 1,1} \right]$. Therefore, ${f_\theta }\left( x \right) \in \left[ {0,1} \right]$, which can be regarded as a membership function.

Here we choose the quantum circuit structure as shown in Fig. \ref{fig_2} as $U\left( {x,\theta } \right)$. This circuit is a quantum circuit that adopts the data re-uploading \cite{perez2020data} method and consists of multiple layers of data encoding circuits and trainable circuit blocks. 

Fig. \ref{fig_2} shows a quantum circuit with only a single qubit, which is composed of multiple layers of circuit blocks, each circuit block containing a data encoding block and a trainable circuit block. Among them, the data encoding block uses the angle encoding method, that is, the input data is encoded into a quantum state by a $R_y$ gate:

\begin{equation}
\label{eq3-1}
\left| {{x_i}} \right\rangle  = {R_y}\left( {{x_i}} \right)\left| 0 \right\rangle ,
\end{equation}
where

\begin{equation}
\label{eq3-2}
{R_y}\left( \theta  \right) = \left[ {\begin{array}{*{20}{c}}
{\cos \frac{\theta }{2}}&{ - \sin \frac{\theta }{2}}\\
{\sin \frac{\theta }{2}}&{\cos \frac{\theta }{2}}
\end{array}} \right]
\end{equation}

It is known that any single qubit gate $U$ (a $2 \times 2$ unitary matrix), which can be discomposed into three rotation gates
\begin{equation}
\label{eq3-3}
U = {e^{i\alpha }}{R_z}\left( \beta  \right){R_y}\left( \gamma  \right){R_z}\left( \delta  \right),
\end{equation}
where, ${e^{i\alpha }}$ is a global phase, $\alpha$ is the global phase factor, $R_z$ is the rotation gate around the Z-axis
\begin{equation}
\label{eq3-4}
{R_z}\left( \theta  \right) = \left[ {\begin{array}{*{20}{c}}
{{e^{ - i\theta /2}}}&0\\
0&{{e^{i\theta /2}}}
\end{array}} \right]
\end{equation}

Since the global phase cannot change the state of a quantum state, any single qubit state $\left| \varphi  \right\rangle $ can be obtained from any initial state $\left| {init} \right\rangle$ by rotating the $R_z, R_y, R_z$ gate at the corresponding angle:
\begin{equation}
\label{eq3-5}
\left| \varphi  \right\rangle  = {R_z}\left( \alpha  \right){R_y}\left( \beta  \right){R_z}\left( \gamma  \right)\left| {init} \right\rangle.
\end{equation}

Then the single qubit quantum circuit can be expressed as
\begin{equation}
\label{eq3-6}
{U_{single\_qubit}} = \prod\limits_l {\left( {{R_z}\left( {\theta _l^3} \right){R_y}\left( {\theta _l^2} \right){R_z}\left( {\theta _l^1} \right){R_y}\left( {{x_i}} \right)} \right)},
\end{equation}
$l$ Indicates the number of layers of the circuit, that is, the number of times the circuit block is repeated. The above quantum circuit only uses a single qubit and does not use a two-qubit gate. Therefore, It is particularly suitable for the current scenario where quantum computing resources are limited.

This layer maps the $k$-th node of input to fuzzy degree.

\begin{equation}
\label{eq4}
{h_{i,k}}^{\left( l \right)} = {f_\theta }^i\left( {{x_k}} \right),\forall i.
\end{equation}
In which, ${h_{i,k}}^{\left( l \right)}$ is the calculation result of the ${i^{th}}$ membership function. After fuzzy membership function processing, we enter the fuzzy rule layer, which is represented by ‘AND’ fuzzy logic, that is ${h_i}^{\left( l \right)} = \prod\nolimits_k {{h_{i,k}}^{\left( {l - 1} \right)}} $. Then, the input is converted into fuzzy degree.

\textbf{Deep neural network representation}. The neural network representation block will obtain the neural representation. Here we generally use a fully connected neural network to extract neural representation of the input. After the ${l^{th}}$ layer of fully connected neural network, we can get

\begin{equation}
\label{eq5}
h{t_i}^{\left( l \right)} = {w_i}^{\left( l \right)}{h^{\left( {l - 1} \right)}} + {b_i}^{\left( l \right)}.
\end{equation}
In which, ${w_i}^{\left( l \right)}$ and ${b_i}^{\left( l \right)}$ represent the weights and bias connecting the ${\left( {l - 1} \right)^{th}}$ layer node to the ${l^{th}}$ layer ${i^{th}}$ node. Then, ReLU activation function will be applied.

\begin{equation}
\label{eq6}
{h_i}^{\left( l \right)} = ReLU\left( {h{t_i}^{\left( l \right)}} \right) = \left\{ \begin{array}{l}
h{t_i}^{\left( l \right)},if\;h{t_i}^{\left( l \right)} > 0\;\\
0\;\;\;,other
\end{array} \right. ,
\end{equation}
${h_i}^{\left( l \right)}$ is the output of the ${l^{th}}$ layer neural network. In order to improve performance, the dropout strategy can be used in each layer of the fully connected network to prevent model over-fitting. It should be pointed out that for different tasks, specific neural network structures can be selected for replacement. For example, for image data, a CNN can be used to replace a fully connected neural network to obtain better classification results.

\textbf{Fusion layer}. The feature fusion is based on the idea of multi-modal fusion \cite{ngiam2011multimodal,deng2016hierarchical}, which merging outputs of multiple neural networks to extract different features of the input data to form neural features, and finally classifying them through the classification layer. We use the same aggregation method as in \cite{deng2016hierarchical} to aggregate the outputs of the neural representation module and quantum fuzzy logic representation. The two representations are responsible for characterizing the fuzzy features and neural features of the data respectively. First, we concatenate the two features and get

\begin{equation}
\label{eq7}
{h_{fus}} = concat\left( {{h_{fuz}},{h_{neu}}} \right).
\end{equation}

Then a fully connected neural network is utilized to fuse the features.

\begin{equation}
\label{eq8}
{h_{fuz}}^{\left( l \right)} = ReLU\left( {{w^{\left( l \right)}}{h_{fuz}}^{\left( {l - 1} \right)} + {b^{\left( l \right)}}} \right).
\end{equation}
In which, ReLU is the activation function, ${w^{\left( l \right)}}$ and ${b^{\left( l \right)}}$ are the weight matrix and bias vector from the ${\left( {l - 1} \right)^{th}}$ layer node to the ${l^{th}}$ layer node respectively. Through multi-layer fully connected network, the output features are deeply fused fuzzy features and neural features.

\textbf{Classifier layer}. The classifier layer processes the fused features and classifies them into corresponding categories. Assuming that there are k categories of data labels in total, through the classifier layer, for the ${i^{th}}$ input sample, a $k$-dimensional output vector will be obtained. Then, a soft-max function is used to calculate the predicted label

\begin{equation}
\label{eq9}
\widehat {{y_i}} = Soft\max (\widetilde {{y_i}}) = \frac{{{e^{\widetilde {{y_i}}}}}}{{\sum\limits_k {{e^{\widetilde {{y_{ik}}}}}} }}.
\end{equation}
In which, $\widehat {{y_i}} = \left[ {\widehat {{y_{i1}}},\widehat {{y_{i2}}},...,\widehat {{y_{ik}}}} \right]$. In this paper, we use the cross-entropy function as the loss function. For $m$ training data, the cross-entropy loss function is calculated as follows

\begin{equation}
\label{eq10}
C = \frac{1}{m}\sum\limits_{i = 1}^m {\sum\limits_{j = 1}^k { - {y_{ij}}\log \widehat {{y_{ij}}} - } \left( {1 - {y_{ij}}} \right)\log \left( {1 - \widehat {{y_{ij}}}} \right)}.
\end{equation}

In the model training phase, the initialization of model parameters and parameter optimization method need to be considered. Parameter initialization is very important in deep neural network learning. A good parameter initialization strategy can help the model converge to a better local minimum point. For the parameters of the classifier layer, we followed the strategy from \cite{glorot2010understanding} for initialization. That is, the bias of each layer in the fully connected neural network is initialized to 0, and the weights are initialized according to the following rules

\begin{equation}
\label{eq11}
{w_{ij}}^{\left( l \right)} \sim U\left[ { - \frac{1}{{\sqrt {{n^{\left( {l - 1} \right)}}} }},\frac{1}{{\sqrt {{n^{\left( {l - 1} \right)}}} }}} \right].
\end{equation}
In which, $U$ is a uniform distribution, and ${n^{\left( {l - 1} \right)}}$ is the number of nodes in the ${\left( {l - 1} \right)^{th}}$ layer. According to different learning tasks, we will fine-tune the value of ${n^{\left( {l - 1} \right)}}$ and do not completely follow the above rules. We follow the Kaiming Uniform method to initialize parameters for the remaining network.

In the parameter optimization part, we first need to calculate the derivative of the loss function for each parameter in the network, and then use the classic optimization algorithm to update the parameters. For the QNN part, i.e., the membership function layer, the gradient can be obtained using the parameter shift method

\begin{equation}
\label{eq12}
\frac{{\partial {f_\theta }\left( x \right)}}{{\partial {\theta _i}}} = \frac{2}{\pi }\left( {{f_{{\theta _i} + \frac{\pi }{2}}}\left( x \right) - {f_{{\theta _i} - \frac{\pi }{2}}}\left( x \right)} \right).
\end{equation}
In which,$\theta  = \left[ {{\theta _1},{\theta _2},...,{\theta _n}} \right]$, ${\theta _i} \pm \frac{\pi }{2} = \left[ {{\theta _1},{\theta _2},...,{\theta _{i \pm \frac{\pi }{2}}},...,{\theta _n}} \right]$. The classic neural network in the model uses the backpropagation algorithm to obtain the gradient of each parameter. Then the SGD algorithm is used to update the model parameters. We set the initial learning rate to $\alpha  = 0.01$, then use the learning rate decay strategy to gradually decrease the learning rate. The decay multiplication factor is set to 0.1. It is proposed that there are many nodes in the neural network layer in the model, and the model can easily be overfitting in training phase. In order to alleviate the phenomenon of over-fitting, we use the dropout strategy, that is, for a certain layer of neural network, some nodes do not participate in the gradient update according to a certain percentage. Dropout probability is set to 0.4. The steps of the entire training are shown in Algorithm \ref{alg:alg1}.

\begin{algorithm}[H]
\caption{HQFNN model training steps.}\label{alg:alg1}
\begin{algorithmic}
\STATE 
\STATE {\textbf{Input:}}training samples and labels $\left( {x,y} \right)$, learning rate $\alpha$, number of categories $k$, hidden layer feature dimension $h$, number of training epochs $N$.
\STATE {\textbf{Components:}}

\STATE Quantum fuzzy logic representation: 
\STATE $FuzMem\left( {Inp\_dim,{\rm{ }}k \times Inp\_dim} \right)$, $Inp\_dim$ indicates the dimension of input;
\STATE Deep neural network representation: 
\STATE $NeurNet\left( {Inp\_dim,{\rm{ }}h} \right)$;
\STATE Fusion layer: $Linear\left( {k,h} \right)$;
\STATE Classifier layer: $ClassifierL\left( {h,{\rm{ }}k} \right)$.

\STATE 
\FOR{$e = 1,...,N$}
\STATE $FuzFea{\rm{ }} = FuzMem\left( x \right)$
\STATE $NeurFea{\rm{ }} = NeurNet\left( x \right)$
\STATE $FuzRulFea{\rm{ }} = {\rm{ }}PROD\left( {FuzFea,\dim  = 1} \right)$
\STATE $FusFea = ADD\left( {Linear\left( {FuzRulFea} \right),NeurFea} \right)$
\STATE $\widehat y = Softmax\left( {ClassifierL\left( {FusFea} \right)} \right)$
\STATE Calculate the loss function C according to $\widehat y$ and $y$;
\STATE Calculating gradients using the backpropagation algorithm and updating all parameters using the SGD algorithm
\ENDFOR
\STATE {\textbf{Output:}}The well-trained model.
\end{algorithmic}
\label{alg1}
\end{algorithm}

\section{Experiments}\label{section:sec4}
\subsection{Experiment Settings}
\subsubsection{Experiment Environment}
The experiment is conducted on the Linux platform. PyTorch is used to construct classical neural networks, while the quantum circuit part is implemented using the Torch Quantum library \cite{wang2022quantumnas}. 

\subsubsection{Datasets}
The experiment chose Dirty-MNIST and 15-Scene data sets to verify the performance of the model.

\begin{figure}[!t]
\centering
\includegraphics[width=3.2in]{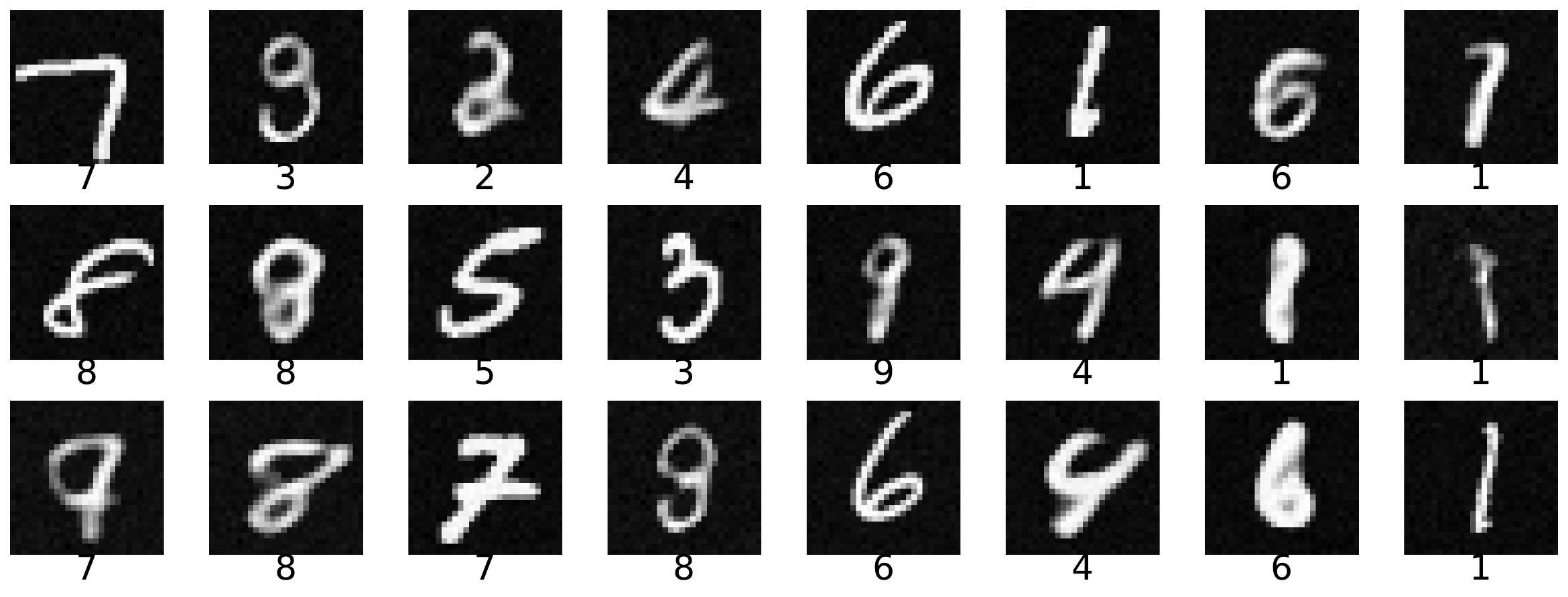}
\caption{Samples from Dirty-MNIST dataset.}
\label{fig_3}
\end{figure}

Dirty-MNIST is a concatenation of MNIST and AmbigouslyMNIST, with 60,000 sample-label pairs in the training set \cite{mukhoti2021deterministic}. AmbigeousMNIST comprises generated ambiguous MNIST samples with varying entropy: 6,000 unique samples, each having 10 labels. By default, the dataset is being normalized and Gaussian noise with stddev 0.05 is added to each sample to prevent acquisition functions from cheating by disgarding "duplicates". Fig. \ref{fig_3} show samples in Dirty-MNIST.

\begin{figure}[!t]
\centering
\includegraphics[width=3.2in]{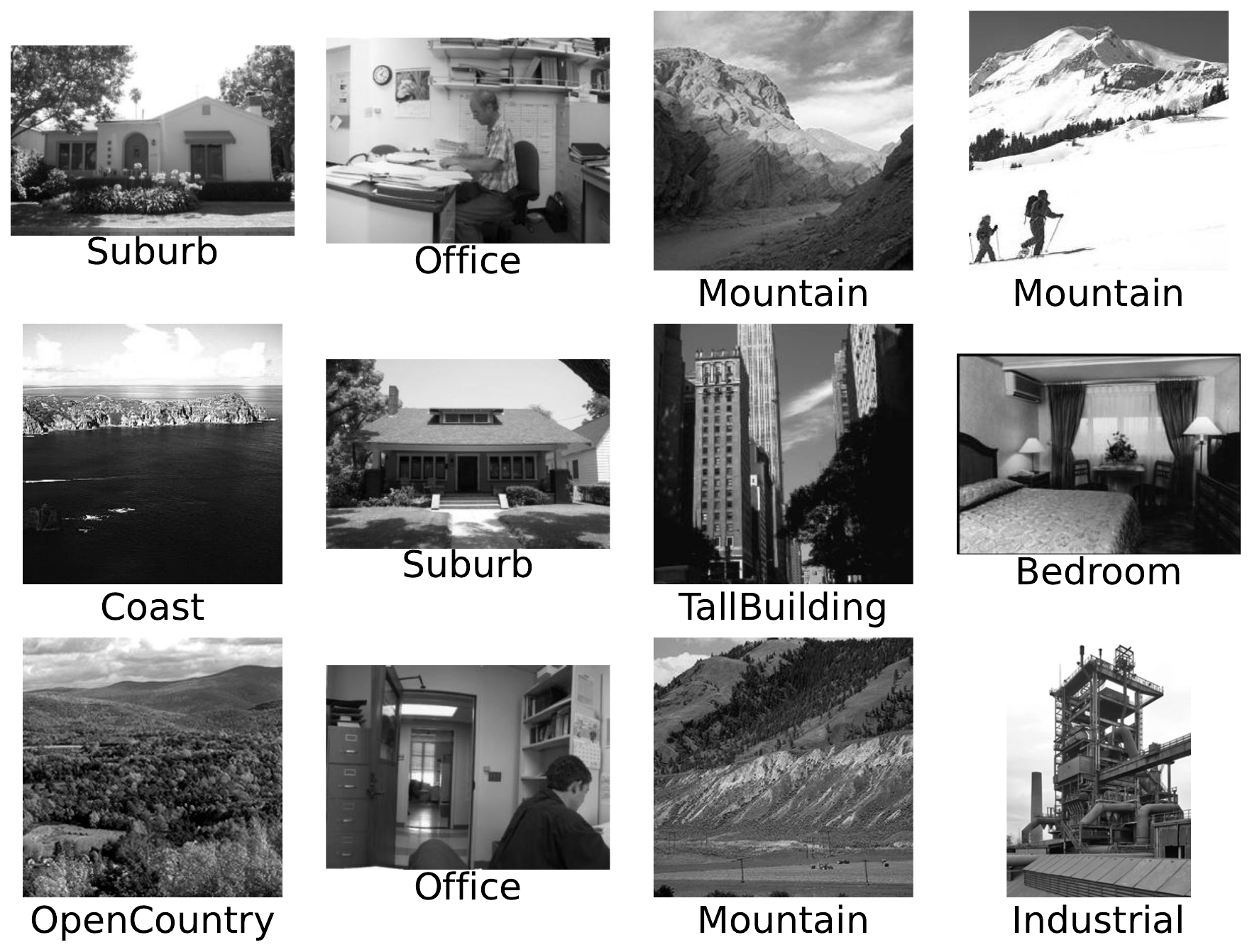}
\caption{Samples from 15-Scene dataset.}
\label{fig_4}
\end{figure}
The 15-Scene dataset includes 15 different natural and urban scenes \cite{scene15}, such as beaches, city streets, forests, offices, kitchens, living rooms, etc. The 15-Scene data set is shown in the Fig. \ref{fig_4}. Since it includes a variety of different types of scenes, from natural landscapes to indoor environments, this dataset provides greater challenges for algorithms, especially in handling the diversity and complexity of the scenes. This data set contains more than 4,500 natural scenery images. We take 100 images from each category as the training set and the remaining images as the test set.

\subsubsection{Hyper-parameters}
The hyper-parameters in this experiment include hyper-parameters in classical neural networks and QNNs. The hyper-parameter settings of the proposed model are shown in TABLE \ref{tab:table1}.

\begin{table}[!t]
\caption{Hyper-parameter settings in the model\label{tab:table1}}
\setlength{\tabcolsep}{12pt}
\setlength\extrarowheight{5pt}
\centering
\begin{tabular}{cc}
\hline
Hyper-parameter & Value\\
\hline
Deep neural network & DNN/CNN\\
Batch size & 128\\
Epochs & 200\\
Qubits & 1\\
Learning rate & 0.01\\
Activation function & ReLU\\
Measurement & Z basis\\
Quantum encoding gate & $R_y$\\
\hline
\end{tabular}
\end{table}

\subsubsection{Evaluation Criteria}
In classification tasks, common metrics for evaluating model performance include accuracy, precision, recall and F1 score. These metrics help us understand how the model performs in different aspects. These metrics are defined as

\begin{equation}
\label{eq13}
Precision = \frac{{TP}}{{TP + FP}}.
\end{equation}

\begin{equation}
\label{eq14}
Recall = \frac{{TP}}{{TP + FN}}.
\end{equation}

\begin{equation}
\label{eq15}
Accuracy = \frac{{TP + TN}}{{TP + TN + FN + FP}}.
\end{equation}

\begin{equation}
\label{eq16}
F1 = \frac{{2Precison \times Recall}}{{Precision + Recall}}.
\end{equation}
TP represents the number of true positive samples; TN represents the number of true negative samples; FP represents the number of false positive samples; FN represents the number of false negative samples.

In multi-class classification tasks, metrics like macro precision, macro recall, and macro F1 are commonly utilized. Macro precision is derived by computing the precision for each class individually and then averaging them. Similarly, macro recall is obtained by averaging the recall across all categories. Finally, macro F1 is calculated from the macro precision and macro recall scores. In a multi-classification task with $k$ categories, macro precision, macro recall, and macro F1 are calculated as follows:

\begin{equation}
\label{eq16-1}
MacroPrecision = \frac{1}{k}\sum\limits_{i \in \left\{ {1,...,k} \right\}} {Precisio{n_i}} ,
\end{equation}

\begin{equation}
\label{eq16-2}
MacroRecall = \frac{1}{k}\sum\limits_{i \in \left\{ {1,...,k} \right\}} {Recal{l_i}} ,
\end{equation}

\begin{equation}
\label{eq16-3}
MacroF1 = \frac{{2MacroPrecision \times MacroRecall}}{{MacroPrecision + MacroRecall}}.
\end{equation}
Among them, $Precision_i$ and $Recall_i$ represent the precision and recall rate of each category respectively. The accuracy calculation method for multi-classification tasks is the same as that for binary classification tasks, that is, the ratio of the number of correctly predicted samples to the total number of samples.

\subsection{Performance Evaluation}

For the Dirty-MNIST dataset, we choose CNN and FDNN as benchmarks to verify the performance of the proposed method. The deep neural network representation part of the proposed model is set to CNN. In order to avoid the influence caused by different deep neural networks, the deep neural network part of FDNN is also set to CNN. The CNN network consists of two layers of convolutional and pooling layers. The convolution kernel size of the first layer is set to $5 \times 5$, with a stride of 1, and it yields 10 output channels. For the second layer, the convolution kernel size is also $5 \times 5$, with a stride of 1, and it yields 20 output channels. Additionally, each pooling layer utilizes a pooling kernel of size 2 with a stride equal to the pooling kernel size. After the processing is completed, the output is flattened into a one-dimensional vector. Subsequently, the data is passed through a fully connected neural network, mapping it to 128-dimensional features. The above-mentioned CNN network is the deep neural network representation part of FDNN or HQFNN. If utilized as an independent CNN model, a classification layer must be added to the network to map the 128-dimensional features to 10-dimensional features, thereby classifying the hidden layer features into respective categories.

\begin{figure}[!t]
\centering
\includegraphics[width=3in]{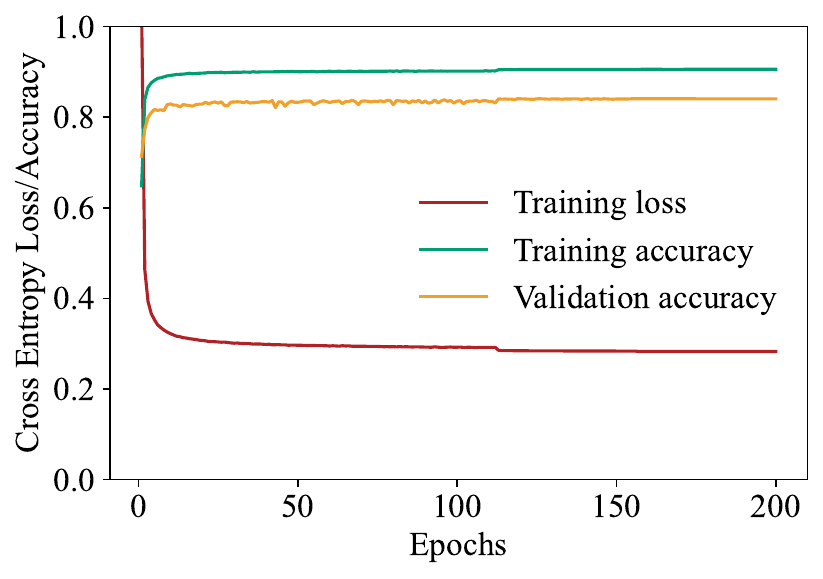}
\caption{The convergence analysis of the training on Dirty-MNIST.}
\label{fig_5}
\end{figure}

In order to observe the convergence of HQFNN, we provide training details as shown in Fig. \ref{fig_5}. It can be seen that in the first few epochs of the model, the loss drops rapidly, and the training accuracy and verification accuracy increase rapidly. At the fifth epoch, the loss dropped to about 0.35, the training accuracy reached about 88.3\%, and the verification accuracy reached about 81.5\%. As the epoch increases, the loss slowly decreases. Until epoch reaches 116, where the learning rate begins to decay, the loss experiences a significant drop. Subsequently, the loss curve gradually flattens out, indicating that the model has converged.

The performance of the proposed model on Dirty-MNIST is shown in TABLE \ref{tab:table2}. It is easy to find that F1 value of the proposed model achieves 85.33\%, which is 2.9\text{\textperthousand} higher than CNN and 1.4\text{\textperthousand} higher than FDNN. It shows that the proposed model can well allow QNNs to participate in training tasks and can achieve good results. In terms of accuracy, the performance of the proposed model is similar to that of CNN and FDNN. Compared with CNN and FDNN, the accuracy of the proposed model is improved by 2.3\text{\textperthousand}, 1.2\text{\textperthousand} respectively. This result shows that compared with FDNN using Gaussian function as fuzzy membership function, the proposed model uses QNN mapping as membership function to improve the accuracy of multi-class classification.

\begin{figure*}[!t]
\centering
\subfloat[CNN]{\includegraphics[width=2in]{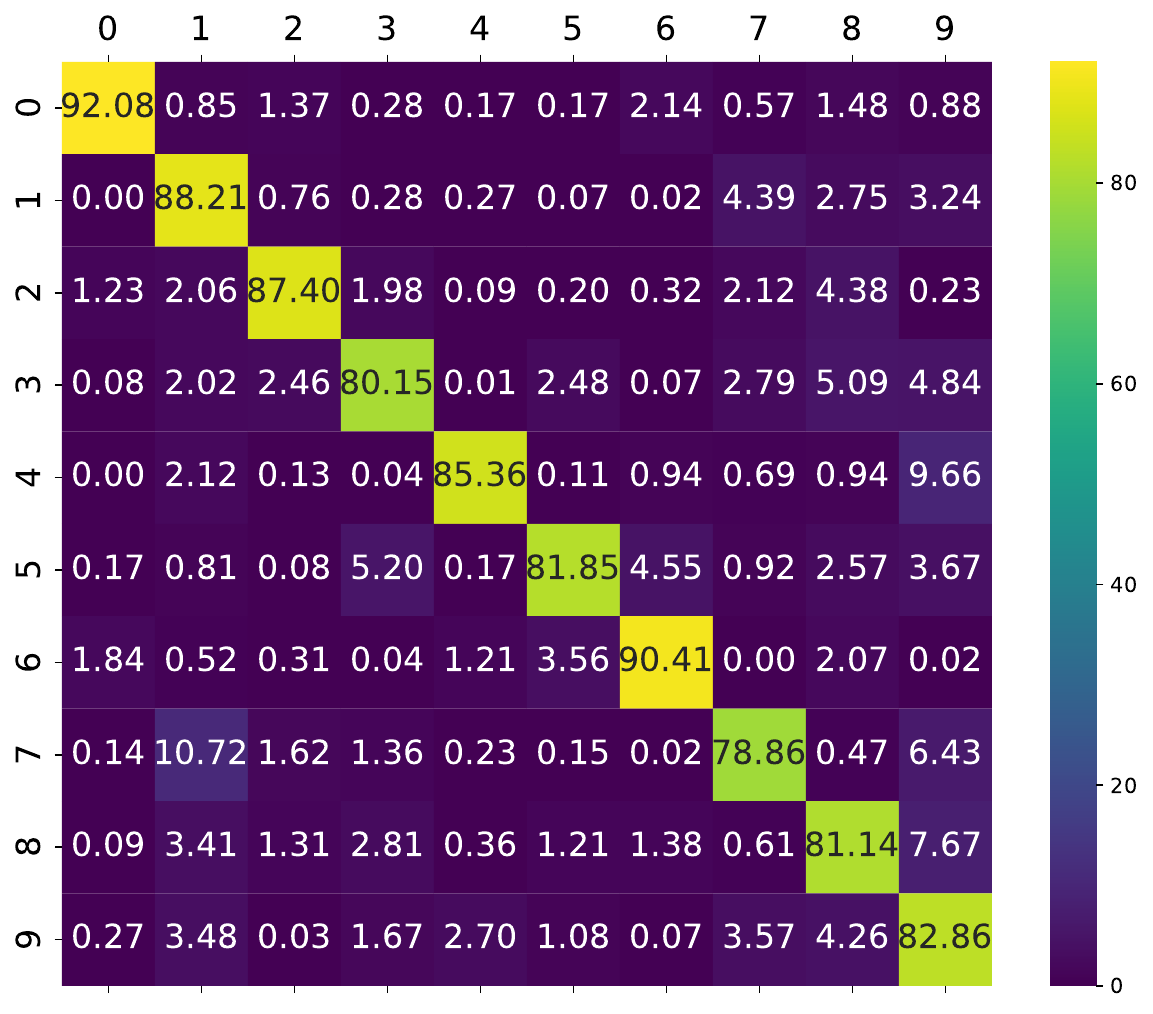}%
\label{DNN0}}
\hfil
\subfloat[FDNN]{\includegraphics[width=2in]{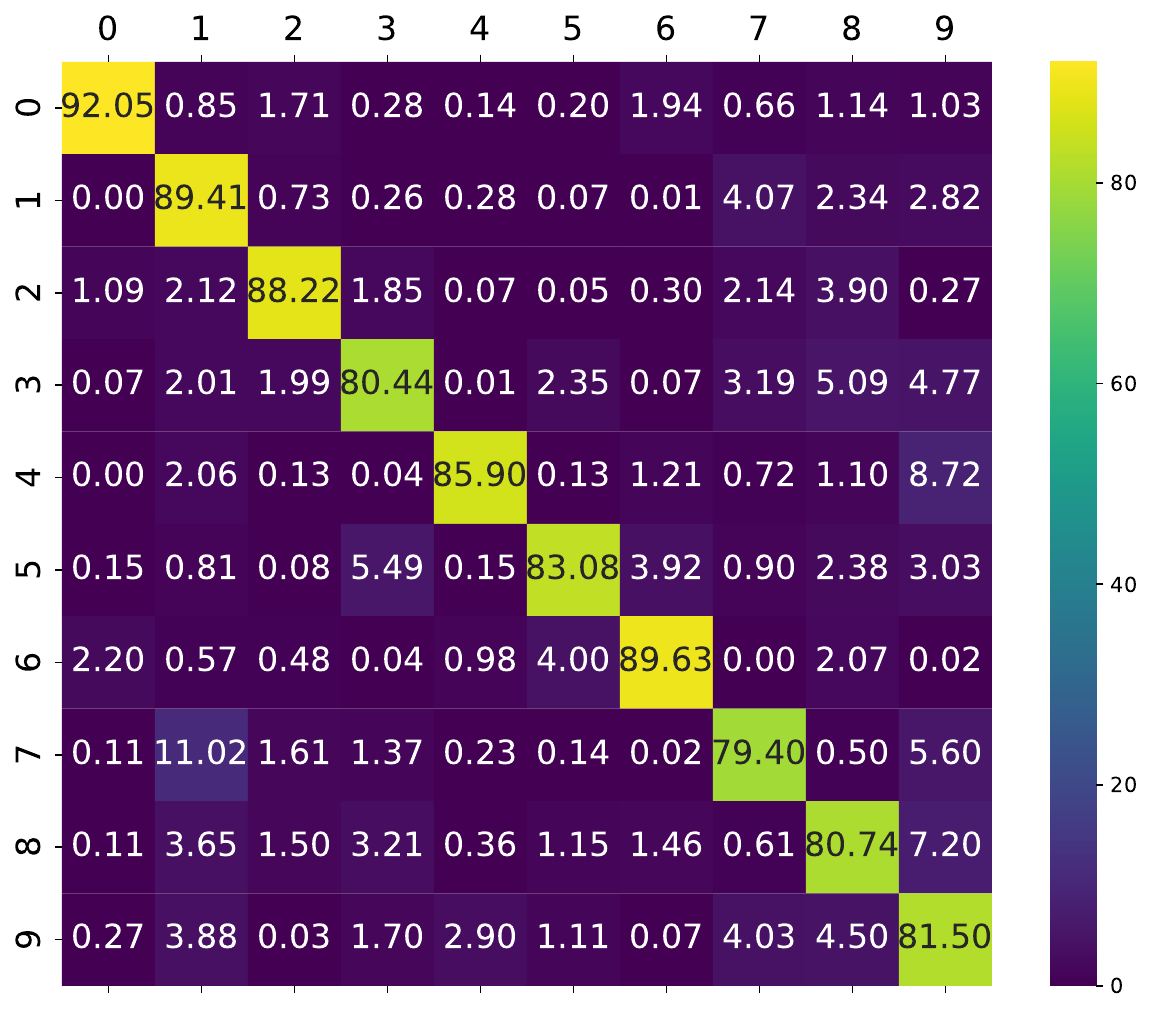}%
\label{FDNN0}}
\hfil
\subfloat[HQFNN]{\includegraphics[width=2in]{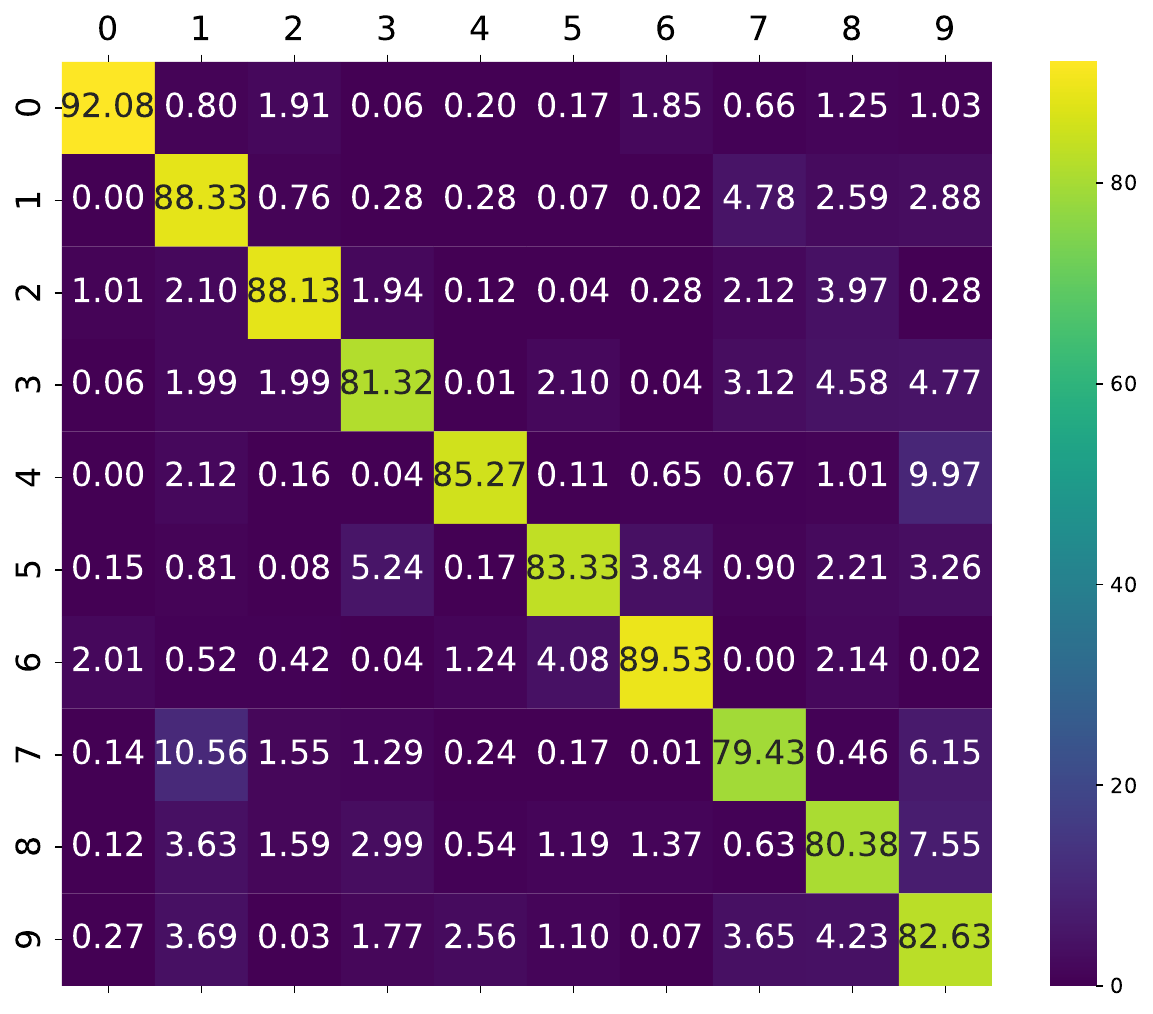}%
\label{HQFNN0}}
\caption{Confusion matrices of different models on Dirty-MNIST.}
\label{confusion_matrix_dmnist}
\end{figure*}

The confusion matrices of the three models on Dirty-MNIST are given in Fig. \ref{confusion_matrix_dmnist}. It can be seen that the three models have higher classification accuracy for '0', '1', '2', and '6', and the classification performance for '7' is the worst. Compared to the other two models, HQFNN exhibits significantly enhanced classification accuracy for the category '3'.

\begin{table}[!t]
\caption{Performance of proposed model on dirty-MNIST\label{tab:table2}}
\setlength{\tabcolsep}{9pt}
\setlength\extrarowheight{5pt}
\centering
\begin{tabular}{ccccc}
\hline
Model & Accuracy & Recall & Precision & F1\\
\hline
CNN & 83.84 & 84.84 & 85.49 & 85.08\\
FDNN & 83.94 & 85.08 & 85.53 & 85.21\\
HQFNN & 84.04 & 85.10 & 85.71 & 85.33\\
\hline
\end{tabular}
\end{table}

\begin{figure}[!t]
\centering
\includegraphics[width=3in]{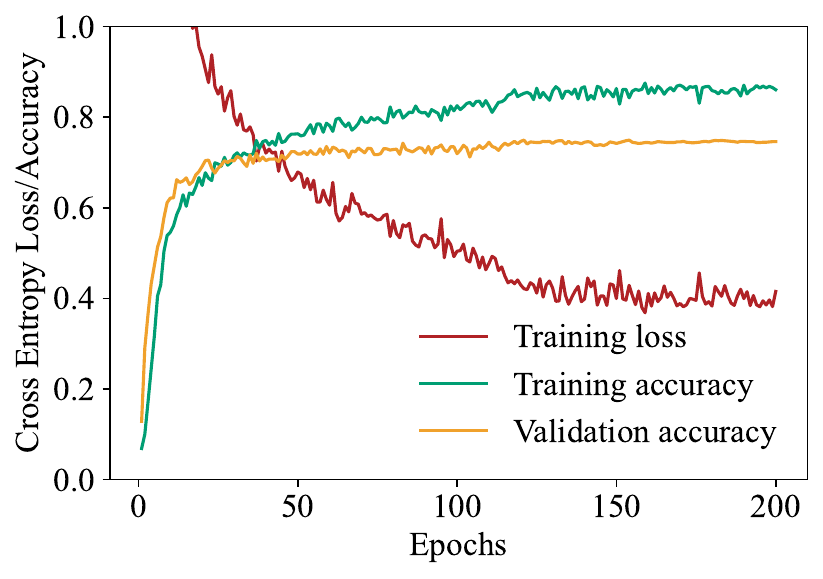}
\caption{The convergence analysis of the training on 15-Scene.}
\label{fig_6}
\end{figure}

On the 15-Scene dataset, we choose DNN and FDNN as benchmarks to evaluate the performance of the proposed model. We first preprocess the 15-Scene data, and we adopt the same method as in \cite{deng2016hierarchical}. Specifically, the dense SIFT features of each image are extracted. Then the local features are clustered to 200 codewords. Finally, the SIFT features are allocated to 200-bit codewords through the core allocation algorithm. We choose a 3-layer $256 \times 256$ fully connected neural network for the hidden layer of DNN. A ReLU activation function is applied to each layer of the network, and the dropout probability is set to 40\%. The deep neural network in the proposed model and FDNN model adopts the same DNN network. The convergence of the proposed model on the 15-Scene dataset is shown in the Fig. \ref{fig_6}. As the epoch increases, it can be seen that the cross entropy of the training set and the verification set gradually decreases, and by 110 epochs, the curve gradually becomes flat; as the epoch further increases, the cross entropy of the training set and the verification set gradually decreases. The cross entropy no longer continues to decrease, indicating that the model has reached convergence.

The performance of the proposed model is shown in TABLE \ref{tab:table3}. It can be seen that the accuracy of the proposed model reaches 74.27\%, which is 3.0\% and 0.6\text{\textperthousand} improved respectively compared to DNN and HDNN. In terms of the value of F1, the proposed model has achieved 74.15\%, which is 3.7\text{\textperthousand} and 1.8\text{\textperthousand} improved compared to DNN and FDNN respectively. On the 15-Scene dataset, the proposed model achieves better performance than DNN and FDNN.

\begin{figure*}[!t]
\centering
\subfloat[DNN]{\includegraphics[width=2.3in]{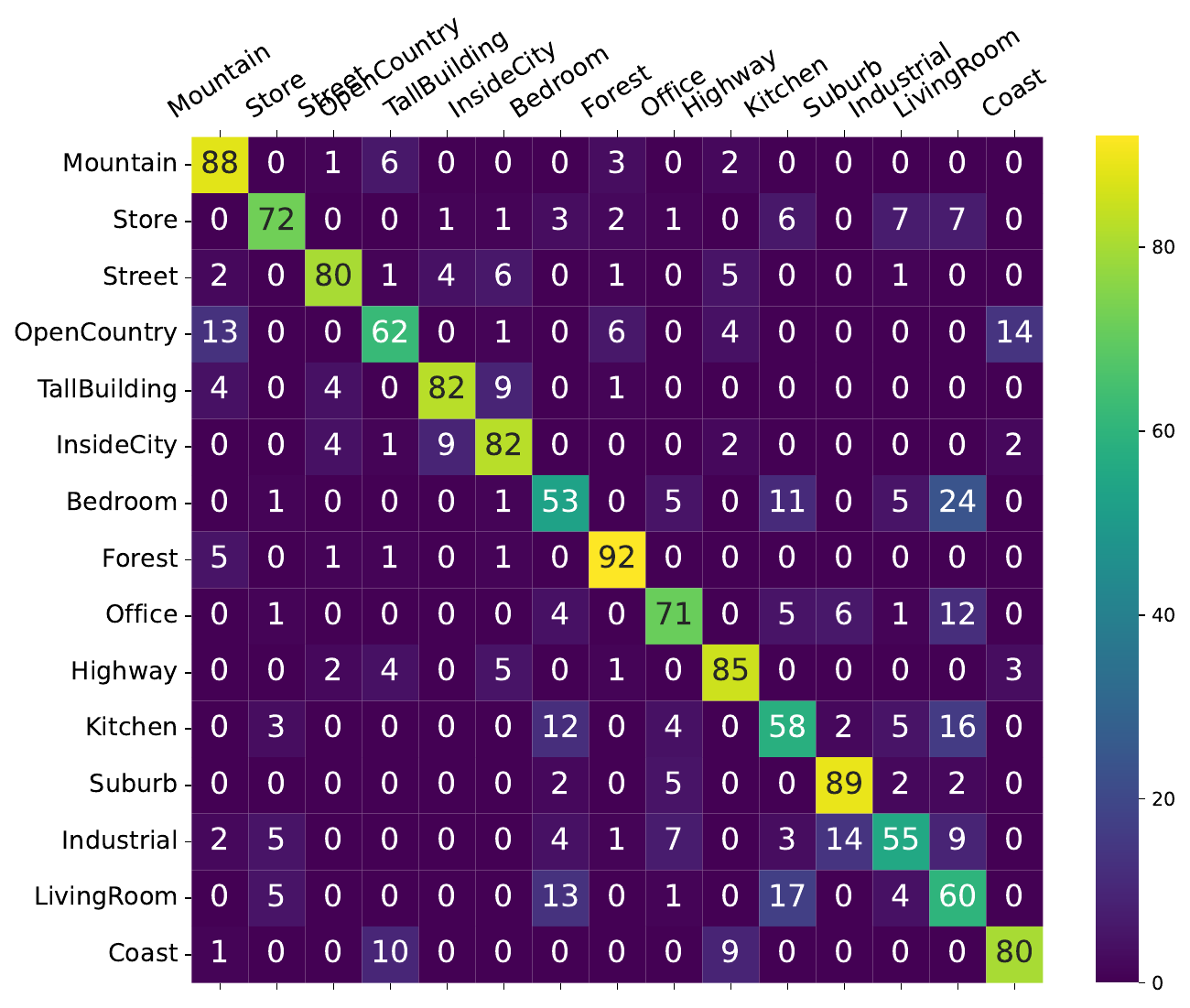}%
\label{DNN}}
\hfil
\subfloat[FDNN]{\includegraphics[width=2.3in]{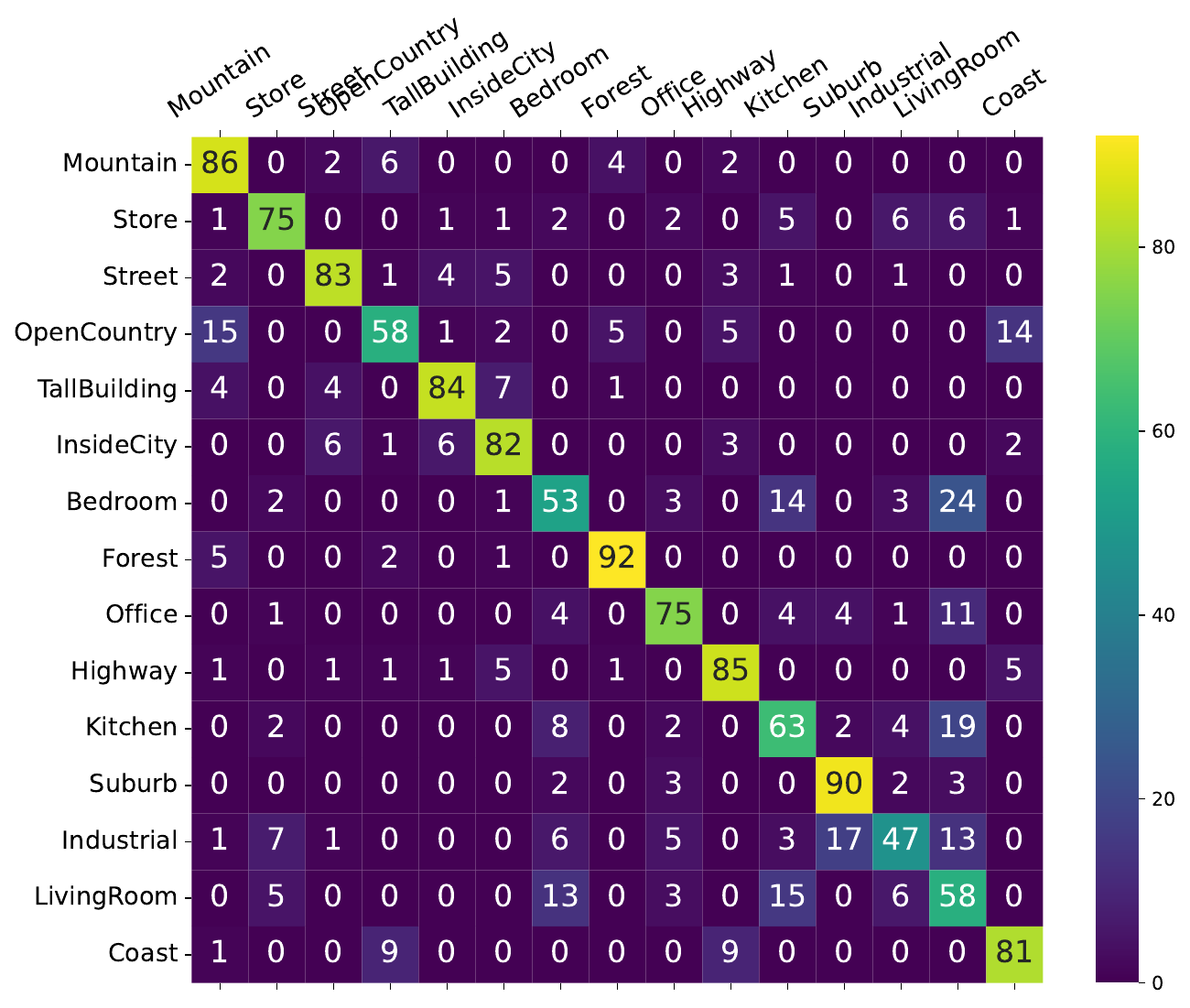}%
\label{FDNN}}
\hfil
\subfloat[HQFNN]{\includegraphics[width=2.3in]{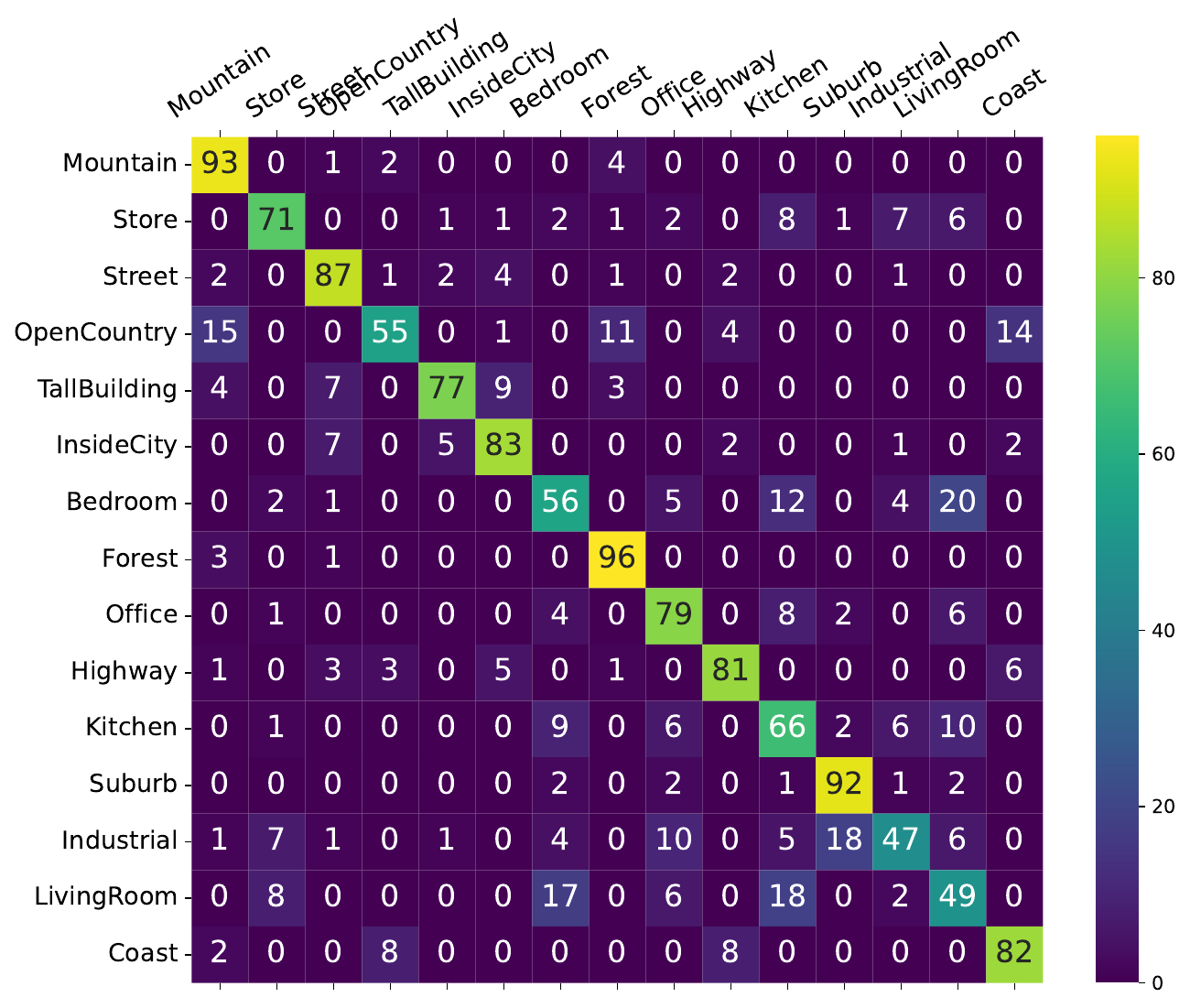}%
\label{HQFNN}}
\caption{Confusion matrices of different models on 15-Scene.}
\label{confusion_matrix_15scene}
\end{figure*}

The confusion matrices of models on the 15-Scene dataset is shown in Fig. \ref{confusion_matrix_15scene}. The three models have high classification accuracy for ‘Mountain’, ‘Forest’, and ‘Suburb’, but poor classification results for ‘Bedroom’, ‘Industrial’ and ‘LivingRoom’. HQFNN has higher classification accuracy for 'Forest' than the other two models.

\begin{table}[!t]
\caption{Performance of proposed model on 15-Scene\label{tab:table3}}
\setlength{\tabcolsep}{9pt}
\setlength\extrarowheight{5pt}
\centering
\begin{tabular}{ccccc}
\hline
Model & Accuracy & Recall & Precision & F1\\
\hline
DNN & 73.99 & 72.13 & 73.06 & 71.96\\
FDNN & 74.13 & 74.13 & 74.62 & 74.10\\
HQFNN & 74.27 & 74.27 & 74.90 & 74.15\\
\hline
\end{tabular}
\end{table}

For fuzzy membership functions, this paper attempts to use QNNs to construct fuzzy membership functions. The proposed QNN uses only single qubit. A layer of QNN contains a ${R_y}$ gate encoding input data and an adjustable parameter quantum gate ${R_z}$, ${R_y}$, ${R_z}$. It is easy to find that a one-layer QNN contains three adjustable parameters. As the number of layers increases, the parameters of this part of the network show a linear growth trend. Compared with the Gaussian membership function, which only has two adjustable parameters, mean and variance. As the number of layers of the QNN increases, the training difficulty of the proposed model also increases, but the advantage is that the QNN can learn more suitable fuzzy membership function.

Regarding that quantum part may be affected by quantum noise, the quantum circuit built by the proposed fuzzy membership function only uses single qubit, does not include two-qubit gates such as the CNOT gate \cite{li2022novel}. Most of the noise in quantum circuits comes from the CNOT gate which requires a long operation time. Therefore, the quantum part is easy to implement and less affected by noise. A detailed discussion of quantum noise will be addressed in the following section.

\begin{figure*}[!t]
\centering
\subfloat[Amplitude damping]{\includegraphics[width=3in]{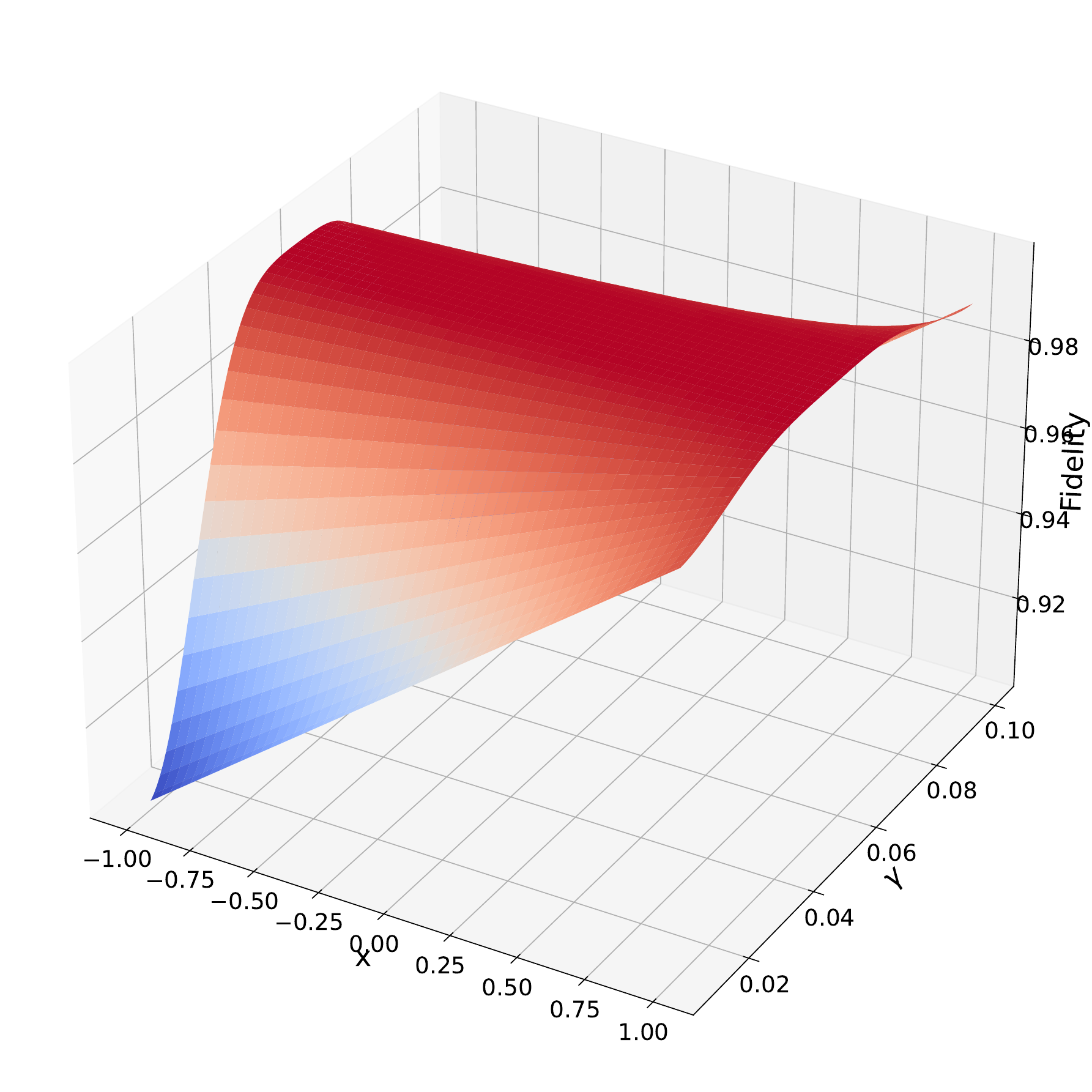}%
\label{AD}}
\hfil
\subfloat[Depolarizing]{\includegraphics[width=3in]{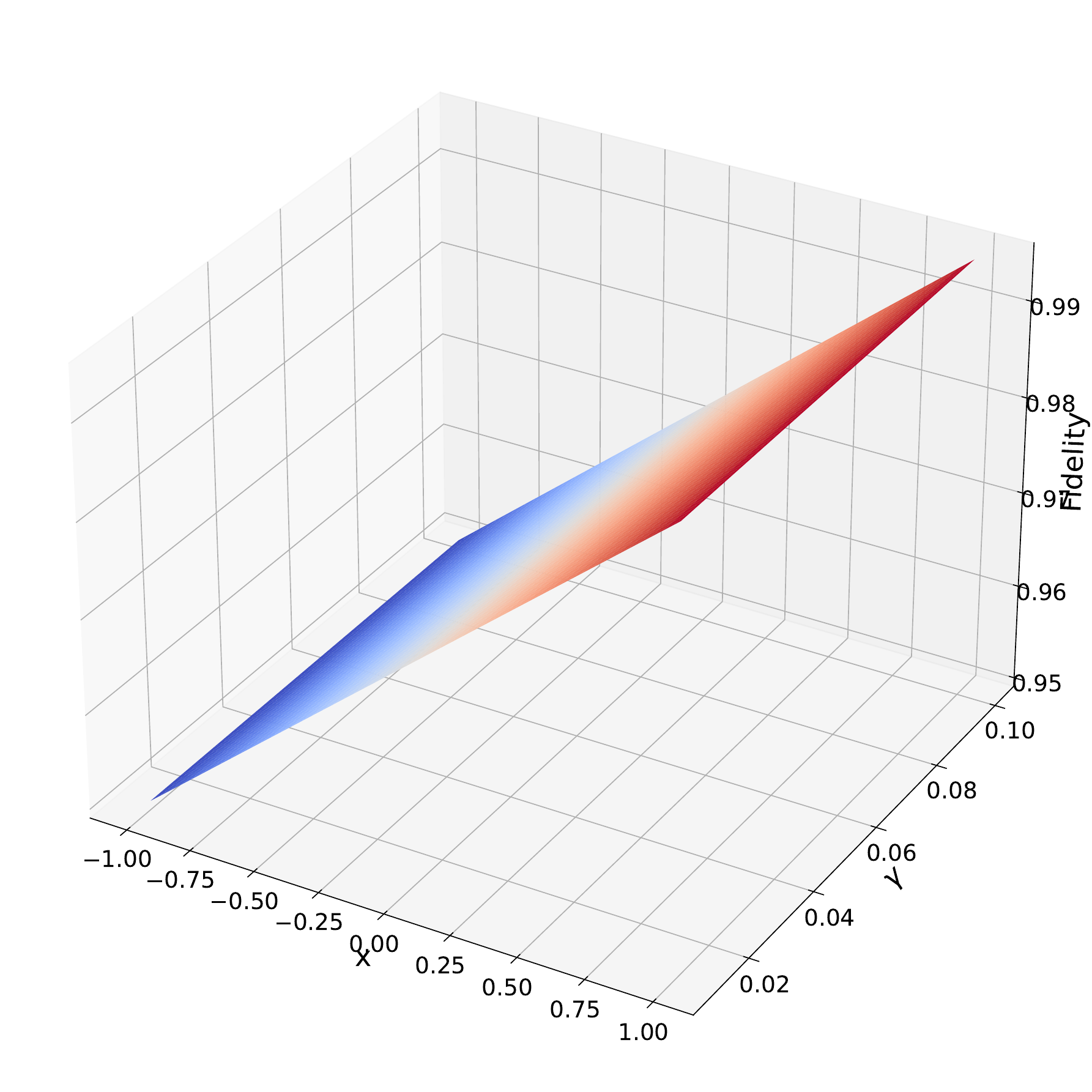}%
\label{DP}}
\caption{The fidelity of output quantum state of quantum circuit changes with input $x$ and noise probability $\gamma$.}
\label{fig_noise}
\end{figure*}

\subsection{Robustness of Quantum Circuits}

During the execution of quantum circuits, the quantum system will be affected by environmental noise. If the quantum circuit has weak resistance to environmental noise, it will seriously affect the accuracy of QNN. Therefore, if environmental noise has a small impact on the quantum circuit, then we say that the quantum circuit has good robustness. In this paper, we mainly analyze quantum circuits with single qubits. We will simulate two types of quantum noise in quantum circuits, which are amplitude damping and depolarization noise:

Amplitude Damping (AD)
\begin{equation}
\label{eq16-4}
{E_0} = \left( {\begin{array}{*{20}{c}}
1&0\\
0&{\sqrt {1 - \gamma } }
\end{array}} \right){\kern 1pt} {\kern 1pt} {\kern 1pt} {\kern 1pt} {\kern 1pt} {\kern 1pt} {\kern 1pt} {\kern 1pt} {E_1} = \left( {\begin{array}{*{20}{c}}
1&{\sqrt \gamma  }\\
0&0
\end{array}} \right)
\end{equation}

Depolarizing (DP)
\begin{equation}
\label{eq16-5}
\begin{array}{l}
{E_0} = \sqrt {1 - \gamma } \left( {\begin{array}{*{20}{c}}
1&0\\
0&1
\end{array}} \right){\kern 1pt} {\kern 1pt} {\kern 1pt} {\kern 1pt} {\kern 1pt} {\kern 1pt} {\kern 1pt} {\kern 1pt} {E_1} = \sqrt {\frac{\gamma }{3}} \left( {\begin{array}{*{20}{c}}
0&1\\
1&0
\end{array}} \right)\\
{E_2} = \sqrt {\frac{\gamma }{3}} \left( {\begin{array}{*{20}{c}}
1&0\\
0&{ - 1}
\end{array}} \right){\kern 1pt} \;\;{\kern 1pt} {\kern 1pt} {\kern 1pt} {\kern 1pt} {\kern 1pt} {\kern 1pt} {\kern 1pt} {E_3} = \sqrt {\frac{\gamma }{3}} \left( {\begin{array}{*{20}{c}}
0&{ - i}\\
i&0
\end{array}} \right)
\end{array}.
\end{equation}
Amplitude damping describes the energy dissipation of quantum systems. Depolarization noise describes the transformation of a quantum state into a maximum mixed state $I/2$ with a certain probability. $\gamma$ is the probability of noise acting on the quantum state.

The evolution process of noise acting on quantum systems is described by operator-sum:
\begin{equation}
\label{eq16-6}
\varepsilon \left( \rho  \right) = \sum\limits_k {{E_k}} \rho {E_k}^\dag ,
\end{equation}
in which, $\left\{ {{E_k}} \right\}$ is called the Kraus operator, satisfies $\sum\limits_k {{E_k}^\dag {E_k} = I}$, and $\varepsilon \left( \rho  \right)$ is the evolved quantum system. We will apply a quantum noise behind each single qubit gate to complete the simulation of noisy quantum circuits.

For quantum circuits containing noise, we use fidelity to quantify the tolerance of quantum circuits to noise. Fidelity represents the degree of similarity of two quantum states:
\begin{equation}
\label{eq16-7}
F\left( {\rho ,\sigma } \right) = Tr{\left( {\sqrt {\sqrt \rho  \sigma \sqrt \rho  } } \right)^2}.
\end{equation}
Here, we calculate the fidelity between the quantum state generated by the quantum circuit without quantum noise and the quantum state generated by the quantum circuit under the influence of quantum noise. The closer the fidelity is to 1, the better the quantum circuit's resistance to noise.

\begin{table}[!t]
\caption{Fidelity of noisy quantum circuit\label{tab:table4}}
\setlength{\tabcolsep}{12pt}
\setlength\extrarowheight{5pt}
\centering
\begin{tabular}{ccc}
\hline
$\gamma$ & AD & DP \\
\hline
0.01 & 0.9964 & 0.9950 \\
0.03 & 0.9894 & 0.9850 \\
0.05 & 0.9823 & 0.9750 \\
0.07 & 0.9751 & 0.9650 \\
0.1 & 0.9644 & 0.9500 \\
\hline
\end{tabular}
\end{table}

The input $x$ of the quantum circuit is a real number in [-1,1]. We sampled 100 input values in [-1,1] and analyzed the change of the noise probability from 0.01 to 0.1 in each input case. The fidelity of output quantum state of quantum circuit changes with input and noise probability as shown in Fig. \ref{fig_noise}. Subsequently, we averaged the fidelity of 100 inputs under the same noise to obtain the quantum circuit fidelity. We show the quantum circuit fidelity when the noise probability is 0.01, 0.03, 0.05, 0.07, and 0.1 in TABLE \ref{tab:table4}. It can be seen that as the noise probability increases, the fidelity of the quantum circuit gradually decreases. When the noise probability reaches 0.1, the fidelity of the quantum circuit can still reach 0.95. Therefore, it can be shown that the proposed HQFNN exhibits robustness against quantum noise.

\begin{figure*}[!t]
\centering
\includegraphics[width=6in]{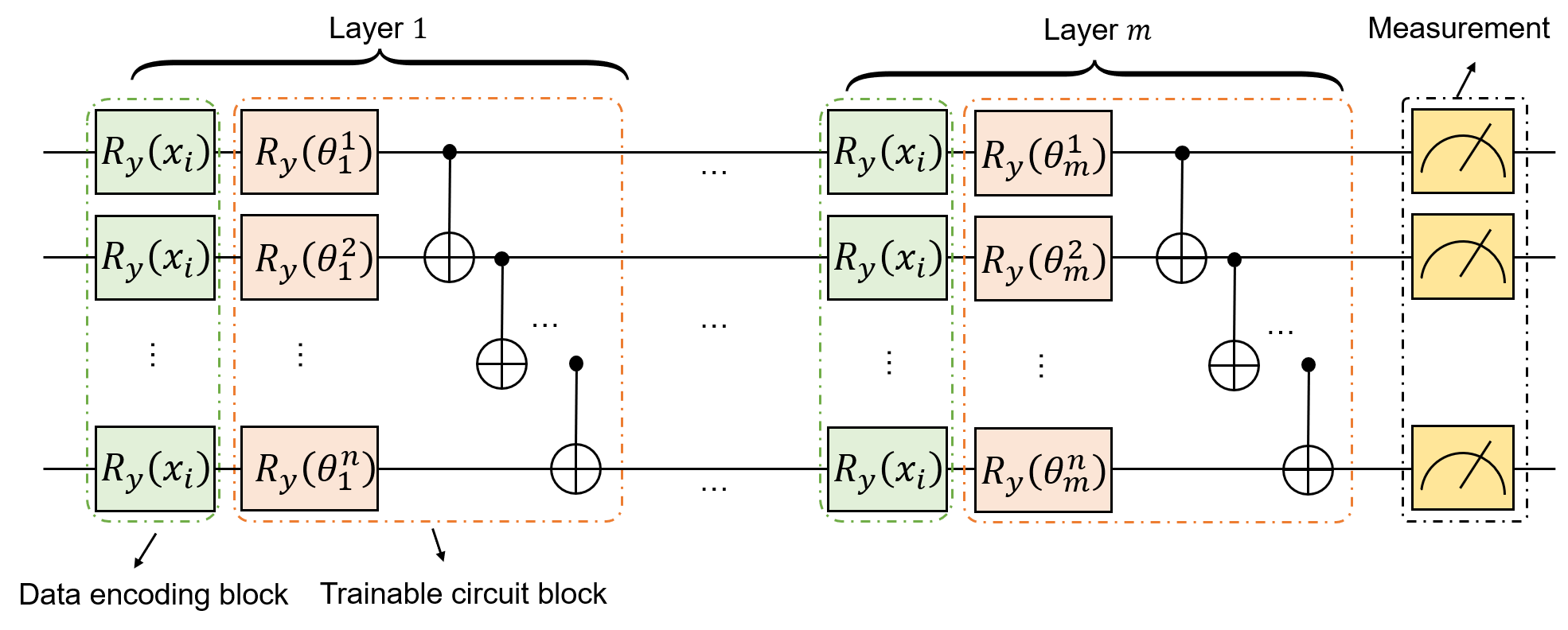}
\caption{Quantum circuit with single-qubit for quantum member function.}
\label{fig_5_2}
\end{figure*}

\section{Discussions}\label{section:sec5}

\subsection{QNNs with More Qubits}
The quantum circuit can be extended to have multiple qubits, we also use multiple layers of circuit blocks which containing encoding blocks and trainable blocks to construct. Assuming that the number of qubits is $n$, in the encoding block we use the $R_y$ gate to encode the element of input to the quantum state.

\begin{equation}
\label{eq3-7}
\left| {{x_i}} \right\rangle  =  \otimes _{q = 1}^n\left( {{R_y}\left( {{x_i}} \right)\left| 0 \right\rangle } \right),
\end{equation}
$\otimes$ indicates Kronecker product. The circuit of the trainable block consists of a rotation layer and an entanglement layer. The rotation layer is composed of $R_y$ gates applied to each qubit 

\begin{equation}
\label{eq3-8}
{U_{Rot}}\left( {{\theta _l}} \right) =  \otimes _{q = 1}^n{R_y}\left( {\theta _q^l} \right),
\end{equation}
where ${\theta _l} = \left[ {\theta _l^1,\theta _l^2,...,\theta _l^n} \right]$. The entanglement layer consists of CNOT gates that applied to two adjacent qubits in sequence 
\begin{equation}
\label{eq3-9}
{U_{Ent}} = \prod\nolimits_{q = 1}^{n - 1} {CNO{T_{\left( {q,q + 1} \right)}}} ,
\end{equation}
where
\begin{equation}
\label{eq3-10}
CNOT = \left[ {\begin{array}{*{20}{c}}
1&0&0&0\\
0&1&0&0\\
0&0&0&1\\
0&0&1&0
\end{array}} \right].
\end{equation}
The subscript of CNOT represents the number of the two qubits it acts on. The first label represents the control qubit, and the second label represents the target qubit. The multi-qubit quantum circuit can be expressed as
\begin{equation}
\label{eq3-11}
{U_{multi\_qubits}} = \prod\limits_l {\left( {{U_{Ent}}{U_{Rot}}\left( {{\theta _l}} \right) \otimes _{q = 1}^n{R_y}\left( {{x_i}} \right)} \right)},
\end{equation}
in which, $l$ represents the number of layers of the circuit. The multi-qubit quantum circuit is shown in Fig. \ref{fig_5_2}. Multi-qubit quantum circuits have better expressive capabilities than single-qubit quantum circuits due to the introduction of more rotation gates and the entanglement introduced by CNOT gates.By constructing multi-qubit QNNs, the model has the potential to achieve higher accuracy. 

\subsection{Data Encoding Methods}
In this paper, we use angle encoding to encode classical data into quantum states, that is, using classical data as the rotation angle of a rotation gate to encode data into the corresponding quantum state. In addition, common encoding methods include amplitude encoding and basis encoding \cite{weigold2020data}. The amplitude encoding method encodes an n-dimensional classical vector using only $\left\lceil {\log n} \right\rceil $ qubits. In HQFNN, the input of the QNN is a real number, so the advantages of amplitude encoding cannot be fully utilized. Basis encoding is to prepare 0 or 1 in binary numbers as $\left| 0 \right\rangle $ or $\left| 1 \right\rangle $ to complete the step of encoding data into the quantum state. The input data in this paper is normalized data, which is a decimal between $\left[ { - 1,1} \right]$. The data first needs to be converted into binary numbers and then encoded into a quantum state. For data with a large number of decimal places, this method requires a very large number of qubits to encode the data. Although more qubits are utilized, this method may effectively capture the differences of classical data, facilitating the extraction of data characteristics. 

\section{Conclusion}\label{section:sec6}
This paper proposes a HQFNN that utilizes QNNs to learn membership functions in classical fuzzy neural network. The proposed model aggregates the fuzzy logic representation extracted by QNNs and the representation of classical neural networks. Simulated experiments are conducted on two image classification tasks, the results show that HQFNN performs better than several existing methods which include FDNN. It is proved that QNNs can be effectively used to learn fuzzy logic representation, which provides ideas for the application of QNN in machine learning. The proposed model has the potential to be used in other machine learning tasks, such as natural language processing, time series prediction, etc.

\section*{Acknowledgments}
This work is supported by National Natural Science Foundation of China (Grant Nos. 62372048, 62371069, 62272056), Beijing Natural Science Foundation (Grant No. 4222031) and the 111 Project B21049.

%

\end{document}